\documentclass[10pt,conference]{IEEEtran}

\IEEEoverridecommandlockouts

\usepackage{cite}
\usepackage{amsmath,amssymb,amsfonts}
\usepackage{algorithmic}
\usepackage{graphicx}
\usepackage{textcomp}
\usepackage{xcolor}
\usepackage{xspace}
\usepackage{tabularx}
\usepackage{hyperref}
\usepackage{balance}
\usepackage{xcolor}
\usepackage{booktabs}
\usepackage[inline]{enumitem}
\usepackage[justification=centering]{caption}
\usepackage{xcolor,colortbl}

\newcommand{\etal}{et al.\xspace}
\newcommand{\ie}{i.e.,\xspace}

\newcommand{\fig}[1]{Fig.~\ref{#1}}
\newcommand{\tab}[1]{Table~\ref{#1}}
\newcommand{\sect}[1]{Section~\ref{#1}}
\newcommand{\changed}[1]{{\color{red}#1}}
\renewcommand\changed[1]{#1}

\newcommand{\github}{GitHub\xspace}
\newcommand{\bitbucket}{BitBucket\xspace}
\newcommand{\gitlab}{GitLab\xspace}

\makeatletter
\newcommand{\linebreakand}{%
  \end{@IEEEauthorhalign}
  \hfill\mbox{}\par
  \mbox{}\hfill\begin{@IEEEauthorhalign}
}
\makeatother

\usepackage[]{todonotes}

\definecolor{myblue}{HTML}{3276AF}
\definecolor{myorange}{HTML}{F38636}

\begin{document}

\title{Identifying bot activity in GitHub pull request and issue comments
\thanks{This research is supported by the Fonds de la Recherche Scientifique -- FNRS under Grant number O.0157.18F-RG43.}
}

\author{\IEEEauthorblockN{Mehdi Golzadeh\IEEEauthorrefmark{1},
Alexandre Decan\IEEEauthorrefmark{1}, Eleni Constantinou\IEEEauthorrefmark{2} and Tom Mens\IEEEauthorrefmark{1}}

\IEEEauthorblockA{\IEEEauthorrefmark{1}Software Engineering Lab,
University of Mons, Belgium\\
Email: \{mehdi.golzadeh,alexandre.decan,tom.mens\}@umons.ac.be}
\IEEEauthorblockA{\IEEEauthorrefmark{2} Eindhoven University of Technology,
Netherlands, 
Email: e.constantinou@tue.nl}}

\maketitle

\begin{abstract}
Development bots are used on Github to automate repetitive activities. Such bots communicate with human actors via issue comments and pull request comments.
Identifying such bot comments allows to prevent bias in socio-technical studies related to software development.
To automate their identification, we propose a classification model based on natural language processing. Starting from a balanced ground-truth dataset of 19,282 PR and issue comments, we encode the comments as vectors using a combination of the bag of words and  TF-IDF techniques.
We train a range of binary classifiers to predict the type of comment (human or bot) based on this vector representation.
A multinomial Naive Bayes classifier provides the best results. Its performance on a test set containing 50\% of the data achieves an average precision, recall, and $F_1$ score of 0.88.
Although the model shows a promising result on the pull request and issue comments, further work is required to generalize the model on other types of activities, like commit messages and code reviews.
\end{abstract}

\section{Introduction}
\label{sec:intro}

Collaborative software development is an inherently social activity carried out on dedicated platforms like GitHub~\cite{Thung2013}.
To automate repetitive tasks carried out by developers, bots are increasingly being adopted by projects to reduce human effort~\cite{Wessel2018}.
Examples of tasks that are automated by bots are refactoring~\cite{Wyrich2019}, dependency updating~\cite{Mirhosseini2017}, generating bug patches~\cite{Monperrus2019}, licence checking and welcoming newcomers.

Bot identification is an essential preprocessing step when analyzing software repositories, especially when human activity is relevant~\cite{DeyMSR2020}.
Recently, researchers have proposed methods  to automatically classify accounts as bot or human like, based on the underlying assumption that bots tend to use more repetitive patterns in their messages than humans.
Dey \etal~\cite{DeyMSR2020} proposed the \textit{BIMAN} method to identify bots that commit code.
\textit{BIMAN} combines three different models that rely (1) on the name of the author, (2) on the content of the commit message, and (3) on files and projects being involved in the commit.
Golzadeh \etal~\cite{GolzadehBotse2020} proposed a classification model to identify bots in \github pull request activity.
Their method measures the similarity of comments and groups them into patterns of similar comments. Bots are then detected based on their lower number of comment patterns.
In a follow-up work, Golzadeh et al.~\cite{Golzadeh2020jss} included more features, considering issue comments as well, and validating the model on a ground-truth dataset composed of several thousands of \github accounts and their associated repositories.

Although these approaches are useful to identify bots at the account level, an account-level classification does not always suffice.
Such a classification can fail in the presence of mixed accounts, i.e., accounts involved in both human and bot activity.
In previous work~\cite{Golzadeh2020jss} we identified 78 accounts out of 5,082 \github accounts combining both activities.
This happens when users grant bots access to post comments on their behalf (e.g., semantic-release bot).
Account-level classifications are too coarse-grained to distinguish human and bot activity at the level of individual comments.
Even in cases where an account is predominantly producing bot (or human) comments, mixed activity may still be observed. For example, human developers may manually produce comments on behalf of a bot when testing this bot in its early stages of adoption.
The above observations call for the need for a more fine-grained classification that is able to identify bot or human activity at the level of individual comments as opposed to the account level.

This paper therefore proposes a classification model to identify bot activity at the level of individual comments. To achieve this, we first transform raw text into machine-understandable features using natural language preprocessing and encoding. Next, we select among a list of machine learning binary classifiers the best performing one in order to classify each comment as bot or human. %
The main value of this approach is the ability to classify comment as originating from a bot or a human without requiring to analyse the entire account's commenting activity. This means that comments can be classified fast, and large datasets can be analysed efficiently.

All the scripts and data used to carry out this study are available in a replication package on:\\
{\small \url{https://doi.org/10.5281/zenodo.4580998}}
%
%
\label{sec:extraction}

In order to train and evaluate a model aiming at distinguishing \github comments created by bots from comments created by humans, we need a large dataset of such pre-labelled comments.
In previous work~\cite{Golzadeh2020jss}, we created a ground-truth dataset of 5,000 accounts that were manually identified by at least two raters as bot or human based on their PR and issue comments. This dataset\footnote{The dataset is available on \url{http://doi.org/10.5281/zenodo.4000388}.} contains 28,287 comments made by 527 bots and 268,504 comments made by 4,473 humans, from which mixed accounts were excluded.

For this study, we extracted from our original dataset~\cite{Golzadeh2020jss} a random, balanced subset of 19,282 comments, composed of 9,641 comments created by 519 bots, and 9,641 comments created by 4,090 humans.

\section{Model construction}
\label{sec:model}

In this section, we propose a machine learning model for classifying \github PR and issue comments.
The preprocessing part of the model combines two widely used techniques in natural language processing for encoding the comment text into machine-understandable features.
The classification part of the model takes as input the preprocessed text to perform the classification task.
In the following paragraphs, we will explain how we constructed the model, which classifier we selected, and how we tuned the parameters to get the best performance and accuracy.

Since human-written texts have no direct meaning for machine learning algorithms, natural language processing is needed to convert such texts into a numerical representation that can be analysed by machines.
A range of different methods can be used to convert raw texts into numerical vectors, such as bag-of-words~\cite{manning2008}, TF-IDF\footnote{Term Frequency - Inverse Document Frequency}~\cite{Sparck1972}, Word2Vec~\cite{Mikolov2013}, and Bert\cite{devlin2019bert}. 

We tested all these preprocessing techniques and their variants and we achieved the highest accuracy with bag-of-words and TF-IDF.
The bag of words technique creates a vector that has as many dimensions as the text corpus has unique words. If a text contains a specific word from the corpus, it will be marked as 1 in the corresponding position of the vector, and 0 otherwise.
TF-IDF is similar except that it assigns a higher weight to both high and low-frequency terms in the document, and the frequency of each term is considered as the indicator of its importance.
Given a comment, we pull out only the unigram words to create an unordered list of words using the bag-of-words method.
Then, TF-IDF is used to form a feature vector, where each feature is a term (\ie word) and the value of the feature is the weight of the term.

For the classification part of our model, we restrict ourselves to binary classifiers since the goal is to classify comments as being produced either by a bot  or a human.
We evaluated a range of binary classifiers: the ZeroR baseline classifier,
Support Vector Machines (SVM)~\cite{soton256459},
multinomial Naive Bayes (NB)~\cite{Lindley1990},
Random Forest (RF)~\cite{breiman2001} %
and k-Nearest Neighbours (kNN)~\cite{Aha1991}.
Since the effectiveness depends on specific input parameters, we followed a standard hyper-parameter tuning process using grid-search cross-validation~\cite{Witten2011}.

We split the comment dataset of \sect{sec:extraction} into a training and a test set (see \tab{tab:traintestsplit}).
The training set will be used as a validation set in a grid-search cross-validation process to determine the best input parameters, the best classifier and to train the selected model. The test set will be used to evaluate the performance of the selected model on new data.
We created both sets so that approximately half of all bot comments (respectively human comments) belong to the training set and the other half belong to the test set.

\begin{table}[!tbh]
    \caption{Number of bot comments and human comments in the training and test set.}
    \label{tab:traintestsplit}
    \centering

    \begin{tabular}{r|rr|r}
        \toprule
        & \bf \# human comments & \bf \# bot comments & \bf total\\
        \midrule
 		\bf training set     &  4,928 & 4,839 & 9,767\\
        \bf test  set          &  4,713 & 4,802 & 9,515\\
          \midrule
	\bf total & 9,641 & 9,641 & 19,282\\
      \bottomrule
    \end{tabular}
\end{table}

While creating both sets, we ensured that comments belonging to the same account were not spread in both sets.
The rationale is that, since comments produced by the same account are more likely to be similar or related, distributing such comments over both sets might lead to unrealistic evaluation results.
More precisely, it could lead the model to be trained on specific words or combinations of words used by a commenter, hence artificially improving the evaluation results if these combinations are also present in the test set.

The performance of the resulting models is measured using the precision $P$, recall $R$ and $F_1$ score for the population of each class (\ie for bot comments $B$ and human comments $H$). We aim to achieve an as high overall $F_1$ score as possible.

We rely on the default parameters for the preprocessing and encoding steps.
We use grid-search cross-validation to find the best classifier and its parameters.
We follow a stratified group k-fold cross-validation process to ensure that each fold preserves the proportion of bot and human comments, and that comments by the same account are not spread across folds.

\tab{tab:gridsearch} reports on the performance for each of the classifiers, in descending order of $F_1$. Only the classifier instances whose parameters resulted in the highest $F_1$ score within each family of classifiers are shown in the table.

\begin{table*}[!h]
    \centering
    \caption{Precision, recall and $F_1$ score of the best classifiers per family of classifiers (in descending order of $F_1$).}
    \label{tab:gridsearch}

    \begin{tabular}{r|cc|cc|ccc}
        \toprule
        & \multicolumn{2}{c|}{\bf bot comments} & \multicolumn{2}{c|}{\bf human comments} & \multicolumn{3}{c}{\bf overall}\\
        \bf classifier & $P(B)$ & $R(B)$ & $P(H)$ & $R(H)$ & $P(B\cup H)$ & $R(B\cup H)$  & $F_1(B\cup H)$ \\
        \midrule
        \bf NB & 0.864 & 0.883 &  0.881 &  0.861 &  0.873 &  0.872 & 0.872 \\
        \bf SVM &  0.971 &  0.718 & 0.777 &  0.979 &  0.874 &  0.848 & 0.845 \\
        \bf RF & 0.898 & 0.542 &  0.672 &  0.935 &  0.785 &  0.739 & 0.727 \\
        \bf KNN &  0.993 &  0.369 & 0.613 &  0.997 &  0.803 &  0.683 & 0.648 \\
        \bf ZeroR & 0.200 & 0.400 & 0.299 &  0.600 &  0.249 &  0.499 & 0.332 \\

        \bottomrule
    \end{tabular}
\end{table*}

We observe that all classifiers substantially outperform the ZeroR baseline classifier; this is a sanity check as this classifier is merely used as a benchmark for other classification methods.
In terms of overall performance, NB and SVM appear to be the most promising classifiers.
The SVM, RF, and KNN classifier have high recall $R(H)$ for human comments (0.979, 0.935 and 0.997, respectively) but a rather low recall $R(B)$ for bot comments (0.718, 0.542 and 0.369, respectively).
The NB classifier, which was obtained using $\alpha= 1.5$ and uniform class prior probabilities%
, has the highest recall for bot comments $R(B)=0.883$ and its overall precision, recall and $F_1$ score is the highest of all considered classifiers.

\section{Model evaluation}
\label{sec:results}

We selected the best classifier NB with the parameters explained in the previous section, and
trained it on the 9,767 comments of the training set.
We evaluated this classification model on the 9,515 new and unseen comments of the test set, of which 4,713 are human comments and 4,802 are bot comments (cf. \tab{tab:traintestsplit}).
 \tab{tab:evalreport} reports the evaluation results.

\begin{table}[h]
    \caption{Evaluation of the Naive Bayes classification model on the test set.}
    \label{tab:evalreport}
    \begin{tabular}{l|ccrrr}
        \toprule
        {}    &  \multicolumn{2}{c}{\bf comments classified as} &  &  & \\
        & \bf bot & \bf human & P & R & $F_1$\\
        \midrule
        \bf bot      &   TP: 4,382 &  FN: 468 &   0.866 &	0.904 & 0.884 \\
        \bf human      &  FP: 680 & TN: 4,172 & 0.900 & 0.860 & 0.880 \\
        \midrule
        \bf average      &   & &    0.882 & 0.882 & 0.882 \\
        \bottomrule
    \end{tabular}
\end{table}

The results show that around 90\% of bot comments (4,382 out of 4,802) and 86\% of human comments (4,172 out of 4,713) are classified correctly.
The overall $F_1$ score is notably good (0.882), indicating that the model generalizes well on unseen data.
The higher recall of bot comments (0.904) and higher precision of human comments (0.900) indicate that the model performs better in detecting bot comments. Nevertheless, we observe a decent $F_1$ score for both classes ($F_1(B) = 0.884$ and $F_1(H)=0.880$), indicating the overall good performance of the model.

\changed{Manual inspection of a sample of misclassified comments revealed that these comments are difficult to classify, even for a human evaluator. For example, the following human comment was misclassified as a bot comment: ``{\em Closing this as resolved by \#72}''. Conversely, the following bot comments were misclassified as human comments: ``{\em %
Here are some suggestions: At index: 33, offset: 6, reason: ``it was'' is wordy or unneeded}''
and
{\em ``If the machine has enough cores, then the work done by the babel loaders can be parallelized to run much faster.''}
}
The selected multinomial Naive Bayes classifier has proven to show a good performance in text classification problems~\cite{Mccallum1998}.
The decision function in this classifier predicts based on the probability computed for each case (\ie each considered comment). If the probability value is above 0.5, then the corresponding case belongs to the target class (in our study, a bot comment) otherwise, it belongs to the complement class (in our study, a human comment).
We gain deeper insight into the predictions made by the model by looking at the associated probabilities. To do so, we extracted the associated probability for test cases.
To some degree, these probabilities correspond to a confidence score: a probability close to 1 indicates that there is a high level of confidence in classifying the comment as bot comment.
Conversely, a probability close to 0 indicates high confidence in classifying the comment as a human comment.
A probability close to 0.5 indicates low confidence in the prediction.

\begin{figure}[h]
	\centering
	\includegraphics[width=0.9\columnwidth]{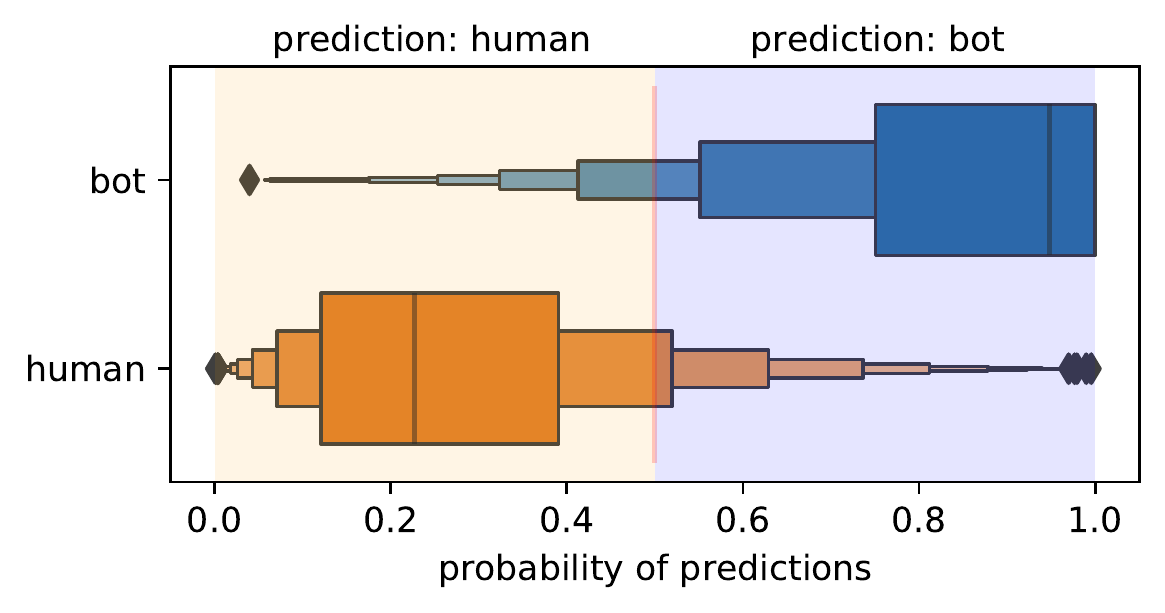}
    \caption{Distribution of the prediction probability for bot comments and human comments}
	\label{fig:confidence}
\end{figure}

\fig{fig:confidence} shows the probability of each prediction, distinguishing between bot and human comments, by means of boxen plots~\cite{Lettervalueplots}.
The misclassified cases correspond to those bot comments for which the probability is below 0.5 (\ie upper left of the figure) and those human comments for which the probability is above 0.5 (\ie lower right of the figure).

We observe that most comments have an associated probability close to 0 for human comments and close to 1 for bot comments, indicating high confidence in the prediction.
Distinguishing between correctly and incorrectly classified cases, we found that bot comments that were correctly classified as such exhibit a median probability of 0.97 (\ie closer to 1 than 0.5, high confidence), while bot comments that were misclassified as human comments have a median probability of 0.37 (\ie closer to 0.5 than 0, low confidence).
A similar observation can be made for human comments: correctly classified human comments have a median probability of 0.20 (\ie closer to 0 than 0.5, high confidence), while the misclassified ones have a median probability of 0.61 (\ie closer to 0.5 than 1, low confidence).
This indicates a higher confidence for the correctly classified comments than for the misclassified ones, suggesting that a probabilistic model is more informative than a simple binary classification model to decide whether a specific comment is produced by a bot or a human.

\section{Discussion}
\label{sec:discussion}

One of the main motivations of this work is to identify comments as having been produced by bots or human. For accounts with mixed activity, such a classification model will be able to split the bot activity from the human activity of the account.
To see whether our model is able to split the bot activity from the human activity of mixed accounts, we applied it on a small dataset provided by a research group from the Eindhoven University of Technology who are currently investigating mixed accounts (see acknowledgements).
Each mixed account comment was labelled as human or bot by at least two researchers of the group, and the dataset consists of 177 bot comments and 203 human comments with full agreement.
Our model was able to correctly classify about 80\% of the cases. Only 3 out of 177 bot comments were misclassified as human comments and 76 out of 203 human comments were misclassified as bot comments.
The model achieved an overall $F_1 = 0.78$ %
with an overall recall $R=0.79$. %
The performance of the model on mixed account comments is promising for our future work, but the identification of human comments in mixed accounts needs to be improved.
One possibility would be to train the model on a dataset that includes mixed accounts as well.
Indeed, the activity of mixed accounts can be different from the one in human-only and bot-only accounts.
However, this would require a substantially larger dataset than the one we have.

In earlier work~\cite{Golzadeh2020jss} we also encountered bots whose comments were partly copied from humans and vice versa.
For example, a translator bot followed a pattern like \emph{``Translation from: $<$translation of some text$>$"} in their comments.
We refer to these comments as mixed comments since they are composed of both human text and bot text.
As a follow-up study, it would be interesting to explore how we can use or adapt the classification model to detect these mixed comments and to extract the human and bot parts of these comments.

The presence of mixed accounts and mixed comments indicates that it is not easy to characterise exactly what a bot comment is. Two different individuals could disagree on the interpretation of comments as being produced by bots.
As a consequence, a more fine-grained classification of (types of) bots would be needed, building further on the work by Erlenhov \etal \cite{Erlenhov2020} who identified three types of bot personas based on their autonomy, chat interface, and smartness.
It would be definitely interesting to explore how the binary classifier we propose can be generalised to detect these different types of bot personas.

\section{Threats to Validity}
\label{sec:threats}

The main threats to validity of this study relate to how \github pull request and issue comments were labeled in the ground-truth dataset of \cite{Golzadeh2020jss}, from which mixed accounts were excluded. The original dataset was constructed by manually labelling \emph{accounts} (as opposed to \emph{individual comments}) as bot or human based on the contents of their comments following an inter-rater agreement process. During this labelling, the raters got to see the 20 most recent comments of each considered account, with an option to see more comments if needed.
The set of comments for this paper was taken from these 20 most recent comments only, in order to increase our confidence that individual comments belonging to an account rated as a bot can actually be regarded as bot comments, and similarly for human comments. %

\changed{One of the limitations is that the proposed classification model is based only on bots that are known today. We cannot claim anything about its performance on new bots in the future. 
Nevertheless, we already observed that the model performed well on the test set, even though 75\% of its vocabulary is new compared to the training set. This suggests that the model can generalize.}

\changed{
A threat to construct validity is that our dataset combines pull request and issue comments of \github accounts, assuming that both types of comments are similar in nature.
To assess whether combining both types of comments affected our results, we built distinct models based on pull request comments and issue comments separately, and we achieved a comparable performance. This suggests that the performance of the classification model is not linked to features associated with the type of comment.

As a threat to external validity, we cannot claim that our model will perform well on other types of comments such as commit messages and code review comments, or on comment data obtained from platforms comparable to GitHub, such as BitBucket or GitLab.
Finally, it is unlikely that our classification model in its present form will perform well on other types of bots, such as chatbots that rely on ever more advanced NLP techniques.}
%
%

%
%
\label{sec:conclusion}

In order to avoid bias due to confounding bots for humans in socio-technical empirical studies of collaborative software development, it is crucial to distinguish human commenting from bot commenting activities.
We proposed a machine learning model to classify individual issue and pull request comments in \github as bot comments or human comments.
To do so, we relied on a dataset of 9,641 human comments and 9,641 bot comments.
The comments were preprocessed using a text vectorisation technique, and different binary classifiers were evaluated on a training set containing roughly half of the data. The best classifier was a multinomial Naive Bayes, with an $F_1$ score of 0.88 on the test set.
When studying the probabilities of the model predictions we found that most misclassified cases are predicted with lower confidence than the correctly classified ones. This suggests that a probabilistic model is more informative than a simple binary classifier.
Further work is still required to address the challenge of mixed accounts, and to generalise the model to other collaborative development platforms such as \gitlab and \bitbucket and other types of activities like commit messages and code review comments.

\section*{Acknowledgement}
We thank Christos Kitsanelis, Nathan Cassee and Alexander Serebrenik from Eindhoven University of Technology for kindly providing the mixed account message dataset.

\newpage
\bibliographystyle{unsrt}
\balance
\bibliography{botsbiblio.bib}

\end{document}